\documentclass[conference]{IEEEtran}
\usepackage{graphicx}
\usepackage{cite}
\usepackage{amsmath,amssymb}
\usepackage[T1]{fontenc}
\usepackage{url}
\usepackage[table]{xcolor}
\newcommand{\best}{\cellcolor{blue!12}}
\newcommand{\stdtiny}[1]{{\scriptsize\textcolor{darkgray}{$\pm$#1}}}  
\IEEEoverridecommandlockouts  

\newif\iffulltext
\fulltexttrue   
\newif\ifsplbuild  

\begin{document}

\title{The Watermark Shortcut: How Provenance Marking Sabotages Audio Deepfake Detection}
\author{%
\IEEEauthorblockN{Nicolas M. Müller\IEEEauthorrefmark{1}\IEEEauthorrefmark{2}, Pascal Debus\IEEEauthorrefmark{1}}
\IEEEauthorblockA{\IEEEauthorrefmark{1}Fraunhofer AISEC, Germany \quad
\IEEEauthorrefmark{2}Resemble AI\\
\{nicolas.mueller, pascal.debus\}@aisec.fraunhofer.de}%
\thanks{This work was supported by the German Federal Ministry of Education and Research (BMBF) under the project AIgenCY (Chancen und Risiken generativer KI in der Cybersicherheit).}}

\maketitle

\begin{abstract}
Provenance watermarking is increasingly treated as a safeguard for synthetic speech, whether built directly into speech-generation models such as Chatterbox, provided through dedicated techniques such as AudioSeal, or deployed by commercial platforms such as ElevenLabs. We identify a previously uncharacterized liability: when synthetic speech is watermarked and human speech is not, detectors trained alongside latch onto the watermark as a spurious ``watermark $\Rightarrow$ fake'' shortcut. This single feature yields three coupled failures: \emph{generalization degradation} (model performance deteriorates on unseen data), \emph{strip-to-evade} (a watermarked fake escapes once unwatermarked), and \emph{mark-to-frame} (watermarking a real voice flags it as fake). In a controlled white-box experiment, a watermark-trained detector shows all three (for example, mark-to-frame lifts Equal Error Rate from $16\%$ to $75\%$). In a black-box test of a commercial API, we show that adding a watermark to real speech disguises it as fake. However, this shortcut is fixable: retraining with the watermark on both classes decorrelates it and restores clean behavior. We release experiment data as a paired clean-versus-watermarked corpus (WASP).

\end{abstract}

\iffulltext
\section{Introduction}

Modern text-to-speech (TTS) systems now synthesize speech that humans can no longer reliably distinguish from genuine recordings~\cite{wang2023neural, chen2024f5tts, du2024cosyvoice}. Such audio deepfakes enable voice-cloning fraud, impersonation, and large-scale disinformation~\cite{sanroman2024proactive, liu2024detecting}. To counter this, the research community builds audio deepfake detectors that separate synthetic (\emph{spoof}) speech from genuine (\emph{bona-fide}) speech~\cite{wang2020asvspoof, jung2022aasist, mueller2022does}.

We identify a previously uncharacterized vulnerability: watermarking synthetic speech for provenance introduces a spurious ``watermark $\Rightarrow$ fake'' feature. For detectors trained on watermarked data, this enables three coupled failures:
\begin{enumerate}
    \item \emph{generalization degradation} (the detector performs worse on ordinary, unwatermarked audio),
    \item \emph{strip-to-evade} (removing the watermark disguises the fake audio as real, i.e. a False Negative), and
    \item \emph{mark-to-frame} (adding a watermark to genuine human speech causes the detector to misclassify it as fake, i.e. a False Positive).
\end{enumerate}
Prior work has noted that watermarking can degrade detector accuracy as a test-time domain shift~\cite{zhang2025impact}. The train-time shortcut, however, and the coupled evasion and framing attacks it enables, have, to our knowledge, not been characterized for speech. We establish this in two regimes, distinguished by the level of access we have to the detector.

\begin{itemize}
  \item \textbf{White-box.} On the ASVspoof19 dataset, we train two detectors that differ only in whether the spoofed training samples are watermarked. The watermark-trained detector (1) generalizes worse to ordinary, unwatermarked audio, (2) misses fakes once their watermark is absent, and (3) frames genuine recordings once a watermark is added; a second detector trained on clean, non-watermarked data exhibits none of these behaviors. This gap to the control pins the cause to the training regime rather than to the watermark as a generic perturbation.
  \item \textbf{Black-box.} Querying a commercial detector through its public API, we reproduce mark-to-frame on genuine speech: adding a watermark to a real recording causes the detector to classify it as more likely fake, despite the absence of any synthesis artifacts.
\end{itemize}

To support the black-box study and enable its replication, we release the corresponding watermarked dataset. Each clip is synthesized with the generator's built-in watermarking disabled and then re-watermarked post hoc, so the clean and watermarked versions of a clip differ only in the watermark.

\begin{table*}[t]
\centering
\caption{White-box experiment: EER (\%) by watermark configuration on the ASVspoof19 evaluation set (top), and external out-of-domain corpora (bottom). A \checkmark marks the watermarked class.
Detectors trained on ASVspoof19, mean$\pm$std reported over three seeds.
The clean-trained detector is not susceptible to either mark-to-frame or strip-to-evade attack, and generalizes better.}
\label{tab:e1}
\footnotesize
\setlength{\tabcolsep}{5pt}
\begin{tabular}{l cc l cc l}
\hline
 & \multicolumn{2}{c}{watermarked} & & \multicolumn{2}{c}{EER (\%)} & \\
\cline{2-3} \cline{5-6}
Test set & bona & spoof & Failure mode & Clean & WM-trained & Interpretation \\
\hline
ASV19 eval  & --          & \checkmark  & baseline       & $10.7$\stdtiny{1.8} & \best$0.7$\stdtiny{0.1} & shortcut, deceptively low EER \\
ASV19 eval  & --          & --          & strip-to-evade & \best$13.8$\stdtiny{1.9} & $31.2$\stdtiny{1.2} & unmarked spoofs raise FNR \\
ASV19 eval  & \checkmark  & --          & mark-to-frame  & \best$16.4$\stdtiny{1.3} & $74.6$\stdtiny{0.9} & FPR \& FNR both rise \\
ASV19 eval  & \checkmark  & \checkmark  & mark-to-frame  & \best$12.9$\stdtiny{1.3} & $17.4$\stdtiny{0.4} & FPR rises, FNR stays low \\
\hline
ASV21-LA    & --          & --          & generalization & \best$14.3$\stdtiny{1.3} & $23.0$\stdtiny{1.3} & worse on unseen fakes \\
In-the-Wild & --          & --          & generalization & \best$18.7$\stdtiny{3.4} & $27.4$\stdtiny{2.8} & worse on unseen fakes \\
\hline
\end{tabular}
\end{table*}

\section{Related Work}

\textbf{Neural speech synthesis and provenance watermarking.} Modern text-to-speech has moved from spectrogram-plus-vocoder pipelines~\cite{shen2018natural} to end-to-end, zero-shot and flow-matching systems~\cite{kim2021conditional, wang2023neural, chen2024f5tts, du2024cosyvoice}. The resulting quality, combined with few-second voice cloning, motivated provenance watermarking by default. For example, Resemble AI's open-source Chatterbox embeds the PerTh watermark in every generated clip \cite{resembleai2025chatterbox}, so a large share of synthetic speech now circulates already watermarked.

\textbf{Audio watermarking, removal, and forgery.} Neural audio watermarking embeds an imperceptible, decodable signal into speech. Resemble AI's PerTh, the watermark shipped by default in Chatterbox, hides such a payload below the auditory perceptual threshold and recovers it with a learned decoder \cite{resembleai2025chatterbox}. Microsoft's WavMark encodes up to 32 bits per second \cite{chen2023wavmark}, Meta's AudioSeal performs sample-level localized detection \cite{sanroman2024proactive}, and Sony's SilentCipher \cite{singh2024silentcipher}, VoiceMark \cite{li2025voicemark}, and WaveVerify \cite{pujari2025waveverify} extend capacity, cloning resistance, and localization. These ship with open implementations, and AudioSeal has watermarked Meta's public audio-generation demos. 
For watermarks to be effective, they need to be hard to remove or forge, but this assumption does not hold. A systematization across 22 schemes and 109 attack configurations finds that none survives all distortions \cite{wen2025sok}, AudioMarkBench shows that AudioSeal, Timbre, and WavMark watermarks fall to no-box, black-box, and white-box perturbations \cite{liu2024audiomarkbench}, and HarmonicAttack achieves near-total removal for most watermarks \cite{li2025harmonicattack}. Passing watermarked audio through a voice-conversion model, even one that re-synthesizes the speaker's own voice, erases the watermark and drives bit extraction to chance across five watermarking systems \cite{ozer2026self}, and overwriting attacks replace a legitimate watermark with a forged one at close to $100\%$ success \cite{yao2025overwriting}. Removability underpins our strip-to-evade attack, and forgeability underpins mark-to-frame. 

The closest prior result to ours studies how watermarks affect detectors directly: Zhang et al.~\cite{zhang2025impact} apply handcrafted and neural watermarks to anti-spoofing data and report that watermarking degrades countermeasure performance, with equal error rate rising as watermark density grows. That work measures a test-time domain shift on a detector trained on clean data and applies watermarks to both classes. We instead study train-time shortcut learning, where the watermark becomes correlated with the spoof label during training, and we show that this single learned feature couples evasion and framing: the same detector can be fooled by stripping a watermark from a fake and by stamping a watermark onto a real voice.

\textbf{Shortcut learning in deepfake detection.} Shortcut learning describes decision rules that succeed on a benchmark but fail under distribution shift, because the model keys on a spurious correlate rather than the intended signal \cite{geirhos2020shortcut}. For example, a classifier trained on dogs photographed outdoors and cats photographed indoors may key on the background rather than the animal, and then misclassify a dog photographed indoors as a cat. The same phenomenon appears in audio anti-spoofing. Chettri et al.~\cite{chettri2020dataset} trace the apparent success of ASVspoof 2017 systems to dataset artifacts in non-speech segments, and M\"uller et al.~\cite{mueller2021speech} show that leading and trailing silence duration alone predicts the label at up to 85\% accuracy, with RawNet2 error rising sharply once silence is trimmed; the silence shortcut recurs in audio-visual datasets \cite{smeu2025circumventing}. Detectors also fail to generalize across generators, with the gap driven by dataset-specific differences rather than newer fakes being harder \cite{mueller2022does, muller2024harder}. A watermark fits this pattern as a spurious feature, and unlike silence, it can be added or removed at will.

\textbf{Watermark and detector interactions.} A few studies examine this watermark--detector intersection directly. In the image domain, Wu et al.~\cite{wu2024watermarks} show provenance watermarks overlap with forgery signals and raise false negatives in passive detectors, and Yu et al.~\cite{yu2025robust} report the same degradation for deep image watermarks. In video, RobustSora~\cite{wang2025robustsora} benchmarks how far detectors depend on the \emph{visible overlay} watermarks that commercial generators add, and shows that augmenting training with de-watermarked and spoofed clips partly recovers accuracy; the audio watermarks we study are instead sub-perceptual perturbations of the signal itself. Guo et al.~\cite{guo2022temporal} and Chen et al.~\cite{chen2025bloodroot} use watermarks as deliberate backdoor triggers in audio models, while AudioMarkNet \cite{zong2025audiomarknet} and FakeMark \cite{ge2025fakemark} intentionally couple a watermark with the fake class for detection or attribution.

\textbf{This work.} These four strands meet here. Synthesis systems mark by default, the marks are removable and forgeable, detectors are known to key on spurious correlates, and the watermark--detector interaction has so far been studied in other modalities, under a malicious trainer, or as a coupling to engineer deliberately. We show that a benign default produces the same coupling as an unintended train-time shortcut in speech, and we quantify the strip-to-evade, mark-to-frame and cross-generator failures it causes together.

\begin{figure}[t]
\centering
\includegraphics[width=0.997\linewidth]{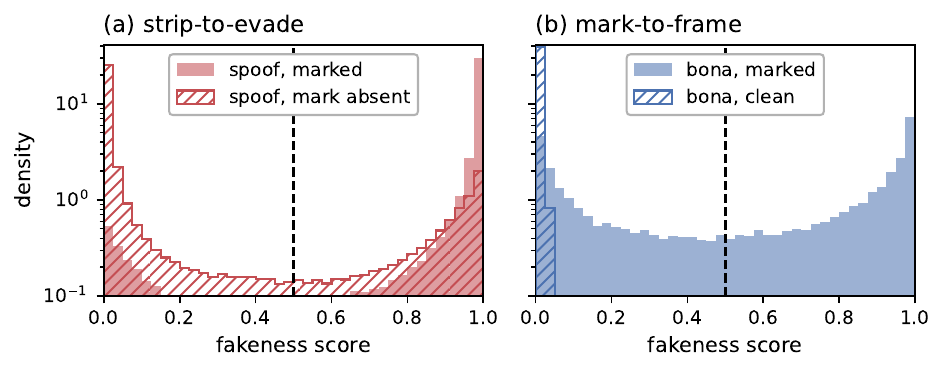}
\caption{Fakeness-score distributions of the watermark-trained detector on the ASVspoof19 evaluation set (three seeds pooled, log density). The dashed line is the $0.5$ decision threshold. Color denotes the class (spoof red, bona-fide blue). A solid fill denotes a watermarked clip and a hatched fill an unwatermarked one. \textbf{(a) strip-to-evade}: removing the watermark from spoofs increases False Negative Rate (incorrect classification as bona-fide) \textbf{(b) mark-to-frame}: adding a watermark to bona-fide instances increases False Positive Rate (incorrect classification as fake). Thus, the watermark alone moves utterances across the decision boundary in both directions.}
\label{fig:dist}
\end{figure}

\section{White-Box Experiment on ASVspoof19}

We aim to show that training on watermarked spoofs introduces the shortcuts described above, and that the effect is \emph{causal}: it arises from the training regime itself rather than from the watermark acting as a generic perturbation at test time.
\ifsplbuild
\begin{table}[t]
\centering
\caption{Dose--response: AASIST retrained with $0\%$, $50\%$, or $100\%$ of training spoofs watermarked (bona fide always clean); mean$\pm$std over three seeds, ASVspoof19 evaluation set.}
\label{tab:rate}
\footnotesize
\begin{tabular}{l ccc}
\hline
Failure mode & $0\%$ & $50\%$ & $100\%$ \\
\hline
Mark-to-frame: FPR (\%)  & $0.3$\stdtiny{0.4}  & $0.2$\stdtiny{0.2}  & $58$\stdtiny{10} \\
Strip-to-evade: FNR (\%) & $37$\stdtiny{9}     & $44$\stdtiny{1}     & $80$\stdtiny{4} \\
Generalization: EER (\%) & $13.8$\stdtiny{1.9} & $19.0$\stdtiny{2.1} & $31.2$\stdtiny{1.2} \\
\hline
\end{tabular}
\end{table}
\fi 

To this end, we train two AASIST~\cite{jung2022aasist} detectors on ASVspoof19 LA~\cite{wang2020asvspoof} under identical settings, differing only in the training data: the clean-trained detector uses the original corpus, while the watermark-trained detector applies the PerTh watermark to every training spoof and leaves bona-fide speech clean. Both are evaluated on the full $71{,}237$-utterance evaluation set under three conditions: no watermarks, evaluation spoofs watermarked, and evaluation bona-fide watermarked. We trim leading and trailing silence, removing the silence-duration shortcut that inflates many leaderboard scores~\cite{mueller2021speech}; our absolute equal error rates (EER) are therefore higher than commonly reported but measure the genuine task rather than a length artifact. The comparison of interest is across columns (clean vs.\ watermark-trained), not against external benchmarks. Performance is reported as mean$\pm$std over three seeds. Taking the spoof class as positive, a false positive (FP) is a genuine clip labeled fake and a false negative (FN) is a missed spoof.

Table~\ref{tab:e1} sweeps the watermark configurations of the ASVspoof19 evaluation set; the clean-trained detector barely moves across any of them, so every effect is attributable to the training regime rather than to the watermark as a test-time perturbation. 
In the real-world baseline, where the spoofs carry the watermark, the watermark-trained detector looks excellent ($0.7\%$ EER), but this is because of the shortcut, not genuine detection capability. 
Strip the watermark off and its false negative rate rises to $80\%$, against the control's $37\%$: \emph{strip-to-evade}. Watermark genuine speech instead and it flags $58\%$ of real recordings as fake, against the control's $0.3\%$: \emph{mark-to-frame}. This framing holds whether or not the surrounding fakes are watermarked (the false positive rate is $58\%$ either way); when the fakes are also watermarked (the Chatterbox default) the EER looks milder ($17.4\%$ vs the control's $12.9\%$) only because those watermarked fakes are still caught, not because there are fewer false positives.
The remaining failure, \emph{generalization degradation}, is the detector's loss on out-of-domain, unwatermarked fakes (last two rows of Table~\ref{tab:e1}): on both ASVspoof2021-LA~\cite{yamagishi2021asvspoof} and In-the-Wild~\cite{mueller2022does}, corpora it never trained on, the watermark-trained model is about $9$ points worse than the control, so the shortcut measurably degrades generalization to real, unwatermarked fakes. 
These fakes carry no watermark to begin with, so this reflects the model, not a removal attack. Figure~\ref{fig:dist} shows the mechanism directly: the watermark alone carries spoofs below the decision threshold and genuine speech above it.

The same pattern appears in two further architectures. Retraining nes2net and TCM-ADD under the identical protocol, EER on watermarked spoofs is $0.3\%$ and $0.5\%$, rising to $20.2\%$ and $21.9\%$ once the watermark is stripped and to $80.3\%$ and $67.0\%$ once genuine speech is watermarked, while their clean-trained controls stay between $5.4\%$ and $10.8\%$ in every condition. The shortcut is therefore not specific to AASIST.

\ifsplbuild
Finally, to probe how much watermark--label correlation the shortcut requires, we retrain AASIST with only $50\%$ of the training spoofs watermarked (Table~\ref{tab:rate}). The failure modes differ: generalization degradation and strip-to-evade grow smoothly with prevalence and are already substantial at $50\%$, whereas mark-to-frame requires the near-total correlation that watermark-on-by-default pipelines produce.
\fi

\section{Black-Box Experiment on a Deployed Detector}

We now ask whether the shortcut is confined to our own controlled detector or already affects systems in the wild. We aim to show that a commercial detector, which we neither trained nor control and access only through its public API, exhibits the same mark-to-frame failure.

\subsection{The WASP corpus}

The black-box study is driven by a paired corpus, WASP, that we build and release.\footnote{\url{https://huggingface.co/datasets/mueller91/WASP}}
We generate speech with six open text-to-speech systems (Fig.~\ref{fig:wasp}) and disable their built-in watermarking. The synthetic audio is therefore free of any watermark. We then pair it with genuine recordings from M-AILABS and AISHELL-3. Finally, each watermark is applied post hoc, namely PerTh, WavMark, AudioSeal, SilentCipher, and a stacked all-watermarks variant, giving for every utterance a matched clean-versus-watermarked pair. The corpus spans English, German, French, and Mandarin. The deployed-detector study below focuses on English; we release the other languages to support further study. Because the watermarks are added after synthesis, any change in a detector's score between the two members of a pair is attributable to the watermark alone, with no confound from the generator. Fig.~\ref{fig:wasp} summarizes its structure.

\begin{figure}[t]
\centering
\includegraphics[width=\linewidth]{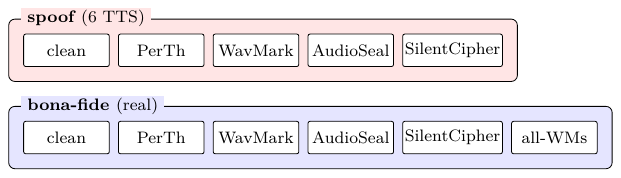}
\caption{The released WASP corpus: every utterance appears clean and under each watermark (all applied post hoc), so the two versions of an utterance differ only in the watermark. Spoofs come from six TTS systems (Chatterbox, Chatterbox-Turbo, DramaBox, Kyutai, Orpheus, Sesame CSM); bona-fide speech is from M-AILABS (en/de/fr) and AISHELL-3 (zh). About $7{,}000$ clips ($\sim$$14$\,h) from $\sim$$1{,}400$ base utterances.}
\label{fig:wasp}
\end{figure}

\begin{table}[t]
\centering
\caption{Mark-to-frame on the deployed commercial detector, applied post hoc to genuine English speech ($n{=}100$ per watermark). We report the detector's mean ``fakeness score'' and the false positive rate (FPR), the fraction of genuine clips it labels fake. The PerTh watermark drives the strongest framing, but SilentCipher and AudioSeal also push genuine speech toward fake; only WavMark leaves the unwatermarked baseline unchanged.}
\label{tab:api}
\footnotesize
\setlength{\tabcolsep}{6pt}
\begin{tabular}{l cc}
\hline
Watermark on bona-fide & Mean fakeness (\%) & FPR (\%) \\
\hline
no watermark & $7.8$  & $4$ \\
PerTh           & $19.3$ & $13$ \\
SilentCipher    & $13.1$ & $10$ \\
AudioSeal       & $13.0$ & $7$ \\
WavMark         & $4.3$  & $0$ \\
all watermarks       & $15.9$ & $8$ \\
\hline
\end{tabular}
\end{table}

\subsection{Attacks on a deployed detector}

On a commercial black-box detector queried through its public API, applying the PerTh watermark to genuine human speech pushes the detector toward calling it fake. 
For real English speech, watermarking raises the classifier's mean ``fakeness score''\footnote{The ``fakeness score'' is an output that the vendor provides for each classification result, where $0\%$ means the model is fully confident the audio is real, and $100\%$ fully confident it is fake.} from $7.8\%$ to $19.3\%$.
Simultaneously, it lowers the true negative rate from $96\%$ to $87\%$, roughly tripling the false positive rate from about $4\%$ to $13\%$. 
The effect is not specific to PerTh (Table~\ref{tab:api}): SilentCipher and AudioSeal also raise the false positive rate above the unwatermarked baseline, as does the stacked all-watermarks variant, and only WavMark leaves it untouched. PerTh drives the strongest framing for a concrete reason the vendor confirmed: its training data contained substantial amounts of Chatterbox, Chatterbox-Turbo, and DramaBox speech, all PerTh-watermarked by default, so the detector learned the PerTh shortcut directly. More broadly, the false positive rate tracks watermark exposure during training: PerTh, dominant in the vendor's data, produces the largest shift; the intermediate shifts under SilentCipher and AudioSeal are consistent with further default-watermarked audio in the training set, and WavMark, which we have no indication the detector trained against, leaves scores at the unwatermarked baseline. The train-time mechanism we isolate in the white-box setting is thus already built into a shipped product.

Strip-to-evade, by contrast, does not transfer. Removing the watermark from a Chatterbox-generated fake, or omitting it altogether, does not prevent detection: across the watermarked and unwatermarked spoofs we queried, the detector flags essentially all of them as fake (mean fakeness near $99\%$, close to $100\%$ detection) regardless of watermark state. Unlike our white-box detector, which is trained only on ASVspoof19 and leans heavily on the watermark, this commercial detector is trained at scale and reads genuine synthesis cues alongside it, so the inherited shortcut frames innocent speech without weakening its detection of real fakes.

\section{Mitigation: Watermark Augmentation}

Since the shortcut resides in the training data, it can be removed at training time. To do so, we retrain the AASIST detector on the union of the clean ASVspoof19 training set and an identical copy in which \emph{every} utterance, bona-fide and spoof alike, carries the PerTh watermark. Each class is thus half watermarked and half unwatermarked, so the watermark is decorrelated from the label and can no longer act as a shortcut. We call this detector \emph{WM-aug} and evaluate it exactly as in the white-box experiment over three seeds.

Table~\ref{tab:mitigation} reports EER on unwatermarked audio, in-domain, and out-of-domain. WM-aug performs similarly to the clean-trained detector everywhere and recovers the generalization capabilities the shortcut detector lost: on ASVspoof2021-LA its EER improves from $23.0\%$ to $14.6\%$, and on In-the-Wild from $27.4\%$ to $17.4\%$, matching the clean-trained detector ($14.3\%$ and $18.7\%$, respectively).

The augmentation also prevents both watermark-based attacks. Because the detector no longer latches onto the watermark, \emph{strip-to-evade} disappears: a spoof is caught with or without the watermark (unwatermarked-spoof EER $31.2 \to 14.4$). Similarly, the \emph{mark-to-frame} attack collapses: stamping a watermark onto genuine speech leaves the false positive rate essentially unchanged, a shift below $1$ percentage point, against the shortcut detector's $58\%$ false positive rate. The watermark becomes inert, which is exactly the intended behavior.

This inertness is what we want from the detector, but it is not a statement about watermarking as a whole.
Where no adversary is present, the presence or absence of a mark can be an informative signal. An audio deepfake detector, however, cannot assume this, since an attacker can add or remove a mark at will. Thus, what we want from a detector is twofold: First, that it does not depend on the watermark, and second, that it performs well by using genuine synthesis artifacts. WM-aug provides both, and leaves provenance verification as a separate channel alongside it.


\begin{table}[t]
\centering
\caption{Mitigation. EER (\%) on unwatermarked audio, both in-domain (ASV19 eval, no watermarks) and out-of-domain, for the clean-trained detector, the watermark-trained detector, and the watermark-augmented detector (WM-aug); mean$\pm$std over three seeds. WM-aug tracks the clean-trained detector and recovers the generalization the shortcut detector loses.}
\label{tab:mitigation}
\footnotesize
\begin{tabular}{l ccc}
\hline
Test set & Clean & WM-trained & WM-aug \\
\hline
ASV19 eval  & $13.8$\stdtiny{1.9} & $31.2$\stdtiny{1.2} & $14.4$\stdtiny{2.4} \\
ASV21-LA    & $14.3$\stdtiny{1.3} & $23.0$\stdtiny{1.3} & $14.6$\stdtiny{1.2} \\
In-the-Wild & $18.7$\stdtiny{3.4} & $27.4$\stdtiny{2.8} & $17.4$\stdtiny{1.3} \\
\hline
\end{tabular}
\end{table}

\section{Conclusion}

Provenance watermarking and the deepfake detectors trained beside it are on a collision course. 
When synthetic speech is watermarked by default and human speech is not, a detector trained on that data learns ``watermark $\Rightarrow$ fake'' rather than genuine synthesis artifacts, and that single spurious feature produces three coupled failures:
\emph{generalization degradation} on ordinary, unwatermarked audio, \emph{strip-to-evade} once a fake's watermark is absent, and \emph{mark-to-frame} once a watermark is added to a real voice.
A clean-trained detector is barely affected by watermarked test data, so the cause is the training regime, not the watermark as a generic perturbation, and the mark-to-frame failure recurs on a deployed commercial detector.
The fix is simple:
applying the same watermarks to both classes during training decorrelates the watermark from the label, which restores the clean-trained detector's behavior and disarms both attacks. 
More broadly, provenance schemes and detectors should be evaluated together rather than developed in isolation. 
We release the paired clean-versus-watermarked corpus to support that work.

\bibliographystyle{IEEEtran}
\bibliography{references}
\fi

\end{document}